\documentclass[12pt,a4paper]{article}
\usepackage[T1]{fontenc}
\usepackage[latin1]{inputenc}
\usepackage{amsmath}

\makeatletter

\providecommand{\LyX}{L\kern-.1667em\lower.25em\hbox{Y}\kern-.125emX\@}

\newcommand{\lyxrightaddress}[1]{
  \par {\raggedleft \begin{tabular}{l}\ignorespaces
  #1
  \end{tabular}
  \vspace{1.4em}
  \par}
}

\newfont{\mcal}{rsfs10 scaled 1200}
\newfont{\bit}{cmbxti10 scaled 1728}

\makeatother

\begin{document}

\vspace{2cm}
{\par\centering {\LARGE Generalized Kerr Schild metrics }\LARGE \par}

{\par\centering {\LARGE and }\LARGE \par}

{\par\centering {\LARGE the gravitational field of a massless particle }\LARGE \par}

{\par\centering {\LARGE on the horizon}\LARGE \par}
\vspace{3.5cm}

{\par\centering {\large Herbert BALASIN\footnote[3]{email:hbalasin@tph.tuwien.ac.at}\footnote[2]{supported by the Austrian science foundation: project P13007-PHY}}\large \par}
\vspace{0.2cm}

{\par\centering \textit{\small Institut für Theoretische Physik}\small \par}

{\par\centering \textit{\small TU-Wien, Wiedner Hauptstraße 8-10,}\small \par}

{\par\centering \textit{\small 1040 WIEN, Austria}\small \par}
\vspace{3cm}

\begin{abstract}
We investigate the structure of the gravitational field generated by a massless
particle moving on the horizon of an arbitrary (stationary) black hole. This
is done by employing the generalized Kerr-Schild class where we take the null
generators of the horizon as the geodetic null vector-field and a scalar function
which is concentrated on the horizon. \vspace{1cm}

\end{abstract}

\lyxrightaddress{\noindent UWThPh--1999-58\\
TUW~~99--20}

\newpage
\section*{\bit Introduction}

Although point-particles play an important role in general relativity as test-masses
following geodesic trajectories, the gravitational field generated by a point-mass
itself is a fairly non-trivial problem. The main difficulty arises from the
incompatibility of (classical) distribution theory with the non-linear character
of the Einstein equations. Nevertheless Aichelburg and Sexl \cite{AS} succeeded
in the construction of the gravitational field of massless particle by taking
the ultrarelativistic limit of the Schwarzschild geometry. The resulting spacetime
is that of an impulsive gravitational wave where the field is completely concentrated
on the null hyperplane of the pulse. The AS-geometry and its higher dimensional
cousins have recently attracted attention as background geometries in string
theory. It also features prominently in \( \text {'t\, Hooft} \)'s work \cite{tH}
on particle scattering at Planckian energies allowing for a closed expression
of the scattering amplitude. Because of the above-mentioned properties of the
AS-geometry there has been quite an active development in possible generalizations.
On the one hand one is looking for ultrarelativistic limits of other black hole
geometries like Reissner-Nordström and Kerr \cite{LouSa,BaNa3,BaNa4,Stein}.
These investigation led in turn to a closer scrutiny of the corresponding ``parent''
geometries, specifically the calculation of distributional energy-momentum tensors
\cite{BaNa1,BaNa2,Ba3}. On the other hand the interpretation of the AS-result
as the gravitational field of a lightlike particle in flat space led to the
question if there exists a generalization describing the gravitational field
of a massless particle in a curved spacetime. \( \text {'t\, Hooft} \) and
Dray \cite{tHD} succeeded in doing so by using Penroses cut and paste approach
\cite{Pen} applied to the Schwarzschild geometry, more precisely to its horizon.
The resulting spacetime describes a particle moving along the horizon and therefore
neccesarily at the speed of light. Due to the singular nature of the spacetime
at the cut and paste surface, their expession for the Ricci tensor unfortunately
contains squares of the delta-function. Following this procedure their work
was generalized by Sfetsos \cite{Sfet} to include cosmological constants as
well as charge. However this generalization still contained the awkward \( \delta ^{2} \)
factors in the expression for the Ricci tensor.

The present paper tries to investigate this situation from the point of view
of the generalized Kerr-Schild class of geometries\footnote[1]{This approach
was actually used by Alsonso and Zamorano \cite{AlZa} in the charged case.
The author is grateful to W. Israel for bringing their work to his attention.}.
It is based on the observation thet both the AS as well as the tHD geomtery
may be interpreted as generalized Kerr-Schild tranforms of their respective
parent geometries, i.e. Minkowski and Schwarzschild. Taking into account that
the gravitational contribution of the particle is completely concentrated on
a null-hypersurface, we investigate the general conditions for a change in the
energy-momnetum tensor corresponding to a massless particle. 

Our work is organized in the following manner: In section one we briefly overview
the generalized Kerr-Schild class and state some of the main results needed
in the following. Section two establishes some properties of ``null Gaussian
coordinates'', whereas scetion three deals with properties of the ``dual''
null-congruence. In section five we use these results to discuss the restrictions
on the null-hypersurface needed to describe a massless particle and derive the
generalized \( \text {'t\, Hooft} \)-Dray equation for the reduced profile.
Finally section five is devoted to show the backward-compatibility of our results
with \cite{tHD} without encountering any undefined \( \delta ^{2} \) expressions.

\section*{\bit 1) Generalized Kerr-Schild class}

This section is intended for a reader who might not be so familiar with the
generalized Kerr-Schild class (GKS) and therefore recapitulates some of its
major properties. A GKS-geometry is characterized by a metric-decomposition

\[
\tilde{g}_{ab}=g_{ab}+fk_{a}k_{b}\qquad k_{a}:=k^{b}g_{ab}\quad \text {and}\quad k^{a}k^{b}g_{ab}=0,\]
 together with the \( g_{ab}- \)geodeticity of \( k^{a} \), i.e. \( (k\nabla )k^{a}=0. \)
As an immediate consequence of the decomposition and the lightlike character
of the vector-field \( k^{a} \) (with respect to \( g_{ab} \) as well as \( \tilde{g}_{ab} \))
one finds
\[
(k\tilde{\nabla })k^{a}=(k\nabla )k^{a},\]
where \( \nabla _{a} \) and \( \tilde{\nabla }_{a} \) denote the Levi-Civita
connections of \( g_{ab} \) and \( \tilde{g}_{ab} \) respectively, which automatically
entails the \( \tilde{g}_{ab}- \)geodeticity of \( k^{a} \) as well. Another
remarkable property of the GKS class of geometries is the \( f- \)linearity
of the mixed Ricci-tensor
\begin{eqnarray*}
 &  & \tilde{R}^{a}\, _{b}=R^{a}\, _{b}-fk^{a}k_{m}R^{m}\, _{b}+\\
 &  & \hspace {2cm}\frac{1}{2}(\nabla _{m}\nabla ^{a}(fk^{m}k_{b})+\nabla ^{m}\nabla _{b}(fk^{a}k_{m})-\nabla ^{2}(fk^{a}k_{b})),
\end{eqnarray*}
or equivalently if one interchanges the covariant derivatives in the expression
within parenthesis
\begin{eqnarray*}
 &  & \tilde{R}^{a}\, _{b}=R^{a}\, _{b}-\frac{f}{2}k^{a}k_{m}R^{m}\, _{b}+\frac{f}{2}R^{a}\, _{m}k^{m}k_{b}+fR^{am}\, _{nb}k_{m}k^{n}\\
 &  & \hspace {1cm}\frac{1}{2}(\nabla ^{a}((f(\nabla k)+(k\nabla )f)k_{b})+\nabla _{b}((f(\nabla k)+(k\nabla )f)k^{a})-\nabla ^{2}(fk^{a}k_{b})).
\end{eqnarray*}
 which is particularly useful if the profile \( f \) is distributional in nature
\cite{Park,BaNa2,Ba3}. Having in mind to apply this decomposition to our situation
of a particle moving along the horizon, it is natural to identify \( k^{a} \)
with the normal of the horizon. This entails in particular that \( k_{a} \)
is hyper-surface orthogonal, i.e. \( \nabla _{a}k_{b}=\nabla _{b}k_{a} \).
Taking a closer look at the last term involving the \( g_{ab}- \)Laplacian
one finds
\begin{eqnarray*}
 &  & \nabla ^{2}k_{b}=\nabla ^{c}\nabla _{b}k_{c}=\nabla _{b}(\nabla k)+R^{m}\, _{b}k_{m},\\
 &  & \nabla _{c}f\nabla ^{c}k^{a}=\nabla _{c}f\nabla ^{a}k^{c}=\nabla ^{a}((k\nabla )f)-(k\nabla )\nabla ^{a}f,\\
 &  & \nabla _{c}k_{b}\nabla ^{c}k^{a}=\nabla _{c}k_{b}\nabla ^{a}k^{c}=-k^{c}\nabla ^{a}\nabla _{c}k_{b}=-(k\nabla )\nabla ^{a}k_{b}+R^{am}\, _{nb}k_{m}k^{n}.
\end{eqnarray*}
Putting everything together the mixed Ricci finally becomes
\begin{eqnarray}
 &  & \tilde{R}^{a}\, _{b}=R^{a}\, _{b}-fk^{a}k_{m}R^{m}\, _{b}+\\
 &  & \hspace {1cm}\frac{1}{2}(\nabla ^{a}((f(\nabla k)+(k\nabla )f)k_{b})+\nabla _{b}((f(\nabla k)+(k\nabla )f)k^{a})-\nonumber \\
 &  & \hspace {1.5cm}\nabla ^{a}(f(\nabla k)+2(k\nabla )f)k_{b}-\nabla _{b}(f(\nabla k)+2(k\nabla )f)k^{a}+\nonumber \\
 &  & \hspace {1.5cm}2(k\nabla )\nabla ^{a}fk_{b}+2(k\nabla )\nabla _{b}fk^{a}+2f(k\nabla )\nabla ^{a}k_{b}+\nonumber \\
 &  & \hspace {1.5cm}\nabla ^{a}f(\nabla k)k_{b}+\nabla _{b}f(\nabla k)k^{a}-\nabla ^{2}fk^{a}k_{b}).\nonumber \label{GKSRic} 
\end{eqnarray}
Although this expression does not look to be much of a gain, we will see in
the following that it actually is.

\section*{\bit 2) Null Gaussian coordinates}

Since we implicitly assumed that the horizon is a particular folium in a family
of null hypersurfaces this section will be concerned with the ``inverse''
question, namely if any given null hypersurface \( \mathcal{N} \) may be embedded
into a family of null hypersurfaces \( \mathcal{N}_{t} \). Actually the question
is quite analogous to the construction of Gaussian normal coordinates. There
one considers a given spacelike hypersurface \( \Sigma  \) and constructs a
family of spacelike hypersurfaces \( \Sigma _{t} \) such that \( \Sigma =\Sigma _{0} \).
Let us take a quick look at this construction. The idea is to use the timelike
unit-normal \( u^{a} \) and extend it geodetically off the surface (i.e. construct
the geodesics starting at \( \Sigma  \) with \( u^{a} \) as tangent vector.)
This provides us with a timelike vector field, whose flow \( \phi _{t} \) allows
us to transport \( \Sigma  \) both into the future and the past, i.e. \( \Sigma _{t}=\phi _{t}(\Sigma ) \).
Now it only remains to show that the folii are orthogonal to \( u^{a} \) and
therefore spacelike hypersurfaces. However since any tangent \( X^{a} \) to
\( \Sigma  \) is Lie-transported along \( u^{a} \) one immediately finds for
the change of the inner product \( u_{a}X^{a} \)
\[
u_{a}(u\nabla )X^{a}=u_{a}(X\nabla )u^{a}=\frac{1}{2}(X\nabla )u^{2}=0,\qquad u^{2}=-1\]
which completes the proof. 

Unfortunately this construction fails in the null case since the normal \( k^{a} \)
of a null hypersurface \( \mathcal{N} \) itself is tangent to \( \mathcal{N} \)
and therefore cannot be used to move off the surface. However, this apparent
drawback has an important consequence, namely the well-known fact that \( \mathcal{N} \)
is ruled by null geodesics generated by \( k^{a} \). Let us now consider an
arbitrary transversal two-surface \( \mathcal{S} \) in \( \mathcal{N} \),
that is a two-surface whose tangent space does nowhere contain \( k^{a} \)
and therefore must be completely spacelike. Transporting \( \mathcal{S} \)
along the flow of \( k^{a} \) we obtain a family of two-surfaces foliating
\( \mathcal{N} \). Obviously those vectors \( \tilde{X}^{a} \) tangential
to \( \mathcal{S} \) remain tangential to \( \mathcal{S} \) upon Lie-transport
along \( k^{a} \), i.e.
\[
k_{a}(k\nabla )\tilde{X}^{a}=k_{a}(\tilde{X}\nabla )k^{a}=\frac{1}{2}(\tilde{X}\nabla )k^{2}=0.\]
This construction immediately provides us with an additional null vector field
\( \bar{k}^{a} \) on \( \mathcal{N} \) defined via
\[
\bar{k}^{2}=0,\, \bar{k}^{a}k_{a}=-1\, \, \text {and}\, \, \bar{k}^{a}\tilde{X}_{a}=0\quad \forall \tilde{X}^{a}\, \, \text {tangential\, to}\, \, \mathcal{S}\]
Extending \( \bar{k}^{a} \) geodetically off \( \mathcal{N} \) and using its
flow \( \phi _{t} \) we may extend \( \mathcal{N} \) into a family of three-surfaces
\( \mathcal{N}_{t}:=\phi _{t}(\mathcal{N}). \) To show that these surfaces
are actually null let us consider the change of the inner product of \( k_{a}\tilde{X}^{a} \)
along the flow, i.e.
\begin{eqnarray*}
(\bar{k}\nabla )k_{a}\tilde{X}^{a}+k_{a}(\bar{k}\nabla )\tilde{X}^{a}=(k\nabla )\bar{k}_{a}\tilde{X}^{a}+k_{a}(\tilde{X}\nabla )\bar{k}^{a}= &  & \\
-\bar{k}_{a}((k\nabla )\tilde{X}^{a}+(\tilde{X}\nabla )k^{a})=2(k\nabla )\bar{k}^{a}\tilde{X}_{a}= &  & \\
-2\bar{k}_{a}(k\nabla )\tilde{X}^{a}=-2\bar{k}_{a}(\tilde{X}\nabla )k^{a}=-2\tilde{X}^{c}(\bar{k}\nabla )k_{c}=-2\tilde{X}^{c}(k\nabla )\bar{k}_{c}=0, &  & 
\end{eqnarray*}
 where the penultimate equality made use of the fact that we may always require
\( \nabla _{a}k_{b} \) to be symmetric at \( \mathcal{N} \) . This shows that
we may always extend a single null hypersurface into a local foliation of hypersurfaces
of the same type.

\section*{\bit 3) Null geodesic congruences}

Since the previous section has established that a null hypersurface \( \mathcal{N} \)
may be (locally) embedded into into a family of null hypersurfaces, let us now
briefly review the properties of the corresponding line congruence, which we
will consider to be affine in addition to being geodetic, i.e. the generating
vector field \( k^{a} \) obeys \( \nabla _{a}k_{b}=\nabla _{b}k_{_{a}} \).
As usual a Jacobi field \( \eta ^{a} \) along a generator of the congruence
is a vector field that is Lie transported along the generator, i.e.
\[
(k\nabla )\eta ^{a}=\eta ^{c}\nabla _{c}k^{a}.\]
Completing \( k^{a} \) to a double null frame \( (k^{a},\bar{k}^{a},m^{a},\bar{m}^{a}) \)
with \( k^{a}\bar{k}_{a}=-1 \), \( m^{a}\bar{m}_{a}=1 \), which we parallel-propagate
along the generator. Decomposition \( \eta ^{a} \) and \( \nabla _{a}k_{b} \)
with respect to this basis and taking into that abreastness (\( k_{a}\eta ^{a}=0 \))
is preserved gives
\begin{eqnarray*}
 &  & \eta ^{a}=\lambda k^{a}+\zeta m^{a}+\bar{\zeta }\bar{m}^{a},\\
 &  & \nabla _{a}k_{b}=\mu k_{a}k_{b}+\rho m_{(a}\bar{m}_{b)}+\alpha k_{(a}m_{b)}+\bar{\alpha }k_{(a}\bar{m}_{b)}+\\
 &  & \hspace {7.5cm}+\sigma m_{a}m_{b}+\bar{\sigma }\bar{m}_{a}\bar{m}_{b}.
\end{eqnarray*}
 Focusing on the \( \zeta  \) equations we obtain
\begin{eqnarray*}
 &  & (k\nabla )\zeta =\frac{\rho }{2}\zeta +\sigma \bar{\zeta },\\
 &  & (k\nabla )\rho +\frac{\rho ^{2}}{2}+2\sigma \bar{\sigma }+R_{ab}k^{a}k^{b}=0,\\
 &  & (k\nabla )\sigma +\rho \sigma +R_{abcd}(k^{a}\bar{m}^{b})(k^{c}\bar{m}^{d})=0,
\end{eqnarray*}
These are the well-known Sachs equations. Since the null-hypersurfaces we are
aiming at are horizons of stationary black holes we may assume that the convergence
\( \rho  \) vanishes. However, the first Sachs equation immediately implies
(by invoking the weak energy-condition) that \( R_{ab}k^{a}k^{b} \) and the
shear \( \sigma  \) must vanish too, which in turn requires \( R_{abcd}(k^{a}\bar{m}^{b})(k^{c}\bar{m}^{d})=C_{abcd}(k^{a}\bar{m}^{b})(k^{c}\bar{m}^{d})=0 \)
telling us that \( k^{a} \) is a principal null-direction of the Weyl-tensor
at \( \mathcal{N} \). Let us now use this results and consider non-abreast
Jacobi-fields, i.e.
\begin{eqnarray*}
 &  & \eta ^{a}=\lambda k^{a}+\zeta m^{a}+\bar{\zeta }\bar{m}^{a}+\omega \bar{k}^{a},\\
 &  & \nabla _{a}k_{b}=\mu k_{a}k_{b}+\alpha k_{(a}m_{b)}+\bar{\alpha }k_{(a}\bar{m}_{b)}
\end{eqnarray*}
which gives the propagation equation
\[
(k\nabla )\lambda =\frac{\bar{\alpha }}{2}\zeta +\frac{\alpha }{2}\bar{\zeta }-\mu \omega ,\qquad (k\nabla )\zeta =-\frac{\alpha }{2}\omega ,\qquad (k\nabla )\omega =0,\]
and via geodetic deviation the change in the propagation coefficients
\begin{eqnarray}
 &  & (k\nabla )\mu +\frac{\alpha \bar{\alpha }}{2}+R_{abcd}k^{a}\bar{k}^{b}k^{c}\bar{k}^{d}=0,\nonumber \\
 &  & (k\nabla )\frac{\alpha }{2}+R_{abcd}\bar{k}^{a}k^{b}k^{c}\bar{m}^{d}=0.\label{nonabreast} 
\end{eqnarray}
Splitting the Riemann tensor into its Weyl and Ricci parts
\[
R^{ab}\, _{bc}=C^{ab}\, _{cd}+2\delta _{[c}^{[a}R_{d]}^{b]}-\frac{1}{6}\delta ^{ab}_{cd}R\]
(\ref{nonabreast}) becomes
\begin{eqnarray*}
 &  & (k\nabla )\frac{\alpha }{2}+C_{abcd}\bar{k}^{a}k^{b}k^{c}\bar{m}^{d}+\frac{1}{2}R_{ab}k^{a}\bar{m}^{b}=0,\\
 &  & (k\nabla )\mu +\frac{\alpha \bar{\alpha }}{2}+C_{abcd}k^{a}\bar{k}^{b}k^{c}\bar{k}^{d}+R_{ab}k^{a}\bar{k}^{b}+\frac{1}{6}R=0.
\end{eqnarray*}

\section*{\bit 4) Impulsive waves on the horizon}

With all the necessary tools assembled let us now return to the main issue of
this work -- impulsive gravitational waves concentrated on the horizon of stationary
black holes. In section two we considered the Ricci tensor of a GKS metric under
the assumption that the null vector field \( k^{a} \) is locally a gradient.
Taking into account the results of the previous section we may take the null
generators of the horizon \( \mathcal{H} \) of a black hole to provide this
vector field. Since we are concerned with the stationary situation the horizon
area can not increase, therefore we have \( \nabla k=0 \) at \( \mathcal{H} \)
. Moreover we will assume that the profile \( f \) is concentrated on \( \mathcal{H} \)
and remains constant along the generators, i.e.\( (k\nabla )f=0. \) This immediately
eliminates the first four terms in the parenthesis of (\ref{GKSRic}). Let us
now consider those terms in the expression for the full Ricci \( \tilde{R}^{a}\, _{b} \)
proportional to \( (k\nabla )\nabla _{b}k^{a} \) and \( R^{a}\, _{b}k^{b} \).
In order to describe a null particle they have to be proportional to the tensor
square of \( k^{a} \) and their coefficients have to be annihilated by \( (k\nabla ) \).
Taking into account \( R_{ab}k^{a}k^{b}=0 \) it follows that 
\[
R^{a}\, _{b}k^{b}=\mu _{r}k^{a}+\alpha _{r}k_{(a}m_{b)}+\bar{\alpha }_{r}k_{(a}\bar{m}_{b)}\]
 and the first requirement becomes \( \alpha _{r}=0 \) or equivalently \( R_{ab}k^{a}\bar{m}^{b}=0 \).
The first condition on the gradient of \( k^{a} \) entails \( (k\nabla )\alpha =0 \),
which together with Ricci-condition requires \( C_{abcd}\bar{k}^{a}k^{b}k^{c}\bar{m}^{d}=0 \).
Since \( C_{abcd}\bar{m}^{a}k^{b}k^{c}\bar{m}^{d}=0 \) we have 
\[
C_{abcd}k^{b}k^{c}=\mu _{w}k_{a}k_{d}+\alpha _{w}k_{(a}m_{b)}+\bar{\alpha }_{w}k_{(a}\bar{m}_{b)}.\]
The constancy of \( \alpha  \) now requires \( \alpha _{w}=0 \) telling us
that \( k^{a} \) has to be a double-null direction of the Weyl-tensor, i.e.
\( C_{abcd}k^{b}k^{c}=\mu _{w}k_{a}k_{d} \) or equivalently \( k_{[p}C_{a]bcd}k^{b}k^{c}=0 \).
The second condition requires \( (k\nabla )^{2}\mu -(k\nabla )\mu _{r}=0, \)
which is equivalent to \( (k\nabla )\mu _{w}-\frac{1}{6}(k\nabla )R=0. \) 

All the terms discussed so far are independent of the profile \( f=\tilde{f}\delta  \).
Let us now take a closer look at those remaining dependent on \( f \) beginning
with the its gradient
\[
(k\nabla )\nabla _{a}f=(k\nabla )(-\tilde{f}(\bar{k}\nabla )\delta k_{a}+\nabla _{a}\tilde{f}\delta )=(k\nabla )\nabla _{a}\tilde{f}\delta .\]
Since we assume that \( \nabla _{a}\tilde{f} \) is a function that is purely
``spatial'' we have
\begin{eqnarray*}
(k\nabla )\nabla _{a}\tilde{f} & = & -k_{a}\bar{k}^{c}k^{d}\nabla _{c}\nabla _{d}\tilde{f}+m_{a}\bar{m}^{c}k^{d}\nabla _{c}\nabla _{d}\tilde{f}+\bar{m}_{a}m^{c}k^{d}\nabla _{c}\nabla _{d}\tilde{f}=\\
 & = & k_{a}\bar{k}^{c}\nabla _{c}k^{d}\nabla _{d}\tilde{f}-m_{a}\bar{m}^{c}\nabla _{c}k^{d}\nabla _{d}\tilde{f}-\bar{m}_{a}m^{c}\nabla _{c}k^{d}\nabla _{d}\tilde{f}=\\
 & = & -k_{a}(\frac{\alpha }{2}(m\nabla )\tilde{f}+\frac{\bar{\alpha }}{2}(\bar{m}\nabla )\tilde{f}).
\end{eqnarray*}
The Laplacian of \( f \) may be decomposed similarly
\[
\nabla ^{2}f=\nabla ^{a}(-\tilde{f}(\bar{k}\nabla )\delta k_{a}+\nabla _{a}\tilde{f}\delta )=\nabla ^{2}\tilde{f}\delta -\tilde{f}(\nabla k)(\bar{k}\nabla )\delta .\]
Focusing on the first term it is convenient to consider the two-dimensional
Laplacian
\begin{eqnarray*}
D^{2}\tilde{f} & = & \sigma ^{ab}D_{a}D_{b}\tilde{f}=\sigma ^{ab}\nabla _{a}D_{b}\tilde{f}=\sigma ^{ab}\nabla _{a}(\sigma ^{c}\, _{b}\nabla _{c}\tilde{f})=\\
 & = & \sigma ^{ab}\nabla _{a}\nabla _{b}\tilde{f}+\sigma ^{ab}\nabla _{a}(k_{b}(\bar{k}\nabla )\tilde{f})=\nabla ^{2}\tilde{f}+2k^{a}\bar{k}^{b}\nabla _{b}\nabla _{a}\tilde{f}=\\
 & = & \nabla ^{2}\tilde{f}-2\bar{k}^{b}\nabla _{b}k^{a}\nabla _{a}\tilde{f}=\nabla ^{2}\tilde{f}+(\alpha (m\nabla )\tilde{f}+\bar{\alpha }(\bar{m}\nabla )\tilde{f}),
\end{eqnarray*}
which finally gives
\[
\nabla ^{2}\tilde{f}=D^{2}\tilde{f}-(\alpha (m\nabla )\tilde{f}+\bar{\alpha }(\bar{m}\nabla )\tilde{f}).\]
Putting everything together the full Ricci-tensor becomes
\begin{eqnarray}
\tilde{R}^{a}\, _{b} & = & R^{a}\, _{b}-\delta \, k^{a}k_{b}\left( \frac{1}{2}D^{2}\tilde{f}+\frac{1}{2}(\alpha (m\nabla )+\bar{\alpha }(\bar{m}\nabla ))\tilde{f}+\right. \nonumber \\
 &  & \hspace {4cm}\left. (\frac{\alpha \bar{\alpha }}{2}-\mu _{w}+\frac{1}{6}R+\frac{1}{2}(\bar{k}\nabla )(\nabla k))\tilde{f}\right) \label{tHDeq} 
\end{eqnarray}
where the concentrated part of the full Ricci is purely two-dimensional, i.e.
all coefficients of the differential operator are annihilated by \( (k\nabla ) \)
if we take into account that \( (\bar{k}\nabla )(\nabla k) \) is annihilated
by \( (k\nabla ) \) on \( \mathcal{N} \). This can be seen from
\[
(k\nabla )\left( (\bar{k}\nabla )(\nabla k)\right) =k^{c}\bar{k}^{d}\nabla _{c}\nabla _{d}(\nabla k)=-\bar{k}^{d}\nabla _{d}k^{c}\nabla _{c}(\nabla k),\]
where the last expression is zero since, due to vanishing of \( \nabla k \)
at \( \mathcal{N} \), \( \nabla _{c}(\nabla k) \) has to be proportional to
\( k_{c} \) at \( \mathcal{N} \).

\section*{\large\bit 5) tHD and AS geometry}

This section is devoted to a short comparison with the results obtained by \( \text {'t\, Hooft} \)
and Dray for a particle moving on the horizon of a Schwarzschild black-hole.
Let us begin with the background geometry in Kruskal coordinates
\begin{eqnarray*}
 &  & ds^{2}=-2A(u,v)dudv+g^{2}(u,v)d\Omega ^{2}\\
 &  & A(u,v)=\frac{m}{r}e^{-r/2m},\quad g(u,v)=r(uv),\\
 &  & e^{r/2m}(\frac{r}{2m}-1)=-\frac{uv}{16m^{2}}.
\end{eqnarray*}
Using an adapted tetrad
\begin{eqnarray*}
e^{u}=du &  & \omega ^{v}\, _{v}=-\frac{\dot{A}}{A}du,\\
e^{v}=Adv &  & \omega ^{i}\, _{u}=\dot{g}\tilde{e}^{i},\, \, \omega ^{i}\, _{v}=\frac{g'}{A}\tilde{e}^{i}\\
e^{i}=g\tilde{e}^{i} &  & \omega ^{i}\, _{j}=\tilde{\omega }^{i}\, _{j},
\end{eqnarray*}
where dot denotes the derivative with respect to \( u \) and prime the derivative
with respect to \( v \), the Riemann tensor at the horizon \( u=0 \) becomes
\begin{eqnarray*}
 &  & R^{uv}=\frac{1}{A}\left( \frac{\dot{A}}{A}\right) 'e^{u}\wedge e^{v}=\frac{1}{4m^{2}}e^{u}\wedge e^{v}\\
 &  & R^{ui}=(\frac{1}{g}\left( \frac{g'}{A}\right) ^{.}+\frac{\dot{A}g'}{gA^{2}})e^{u}\wedge e^{i}+\frac{1}{gA}\left( \frac{g'}{A}\right) 'e^{v}\wedge e^{i}=-\frac{1}{8m^{2}}e^{u}\wedge e^{i}\\
 &  & R^{vi}=\frac{1}{g}\left( \ddot{g}-\frac{\dot{g}\dot{A}}{A}\right) e^{u}\wedge e^{i}-\frac{\dot{g}'}{gA}e^{v}\wedge e^{i}=-\frac{1}{8m^{2}}e^{v}\wedge e^{i}\\
 &  & R^{ij}=\frac{1}{g^{2}}\left( 1+\frac{2\dot{g}g'}{A}\right) e^{i}\wedge e^{j}=\frac{1}{4m^{2}}e^{i}\wedge e^{j}.
\end{eqnarray*}
 From this expression it is to read off \( \mu _{w}=1/4m^{2}. \) Since the
generators of the horizon coincide with \( -du \), we have \( k^{a}=(1/A)\partial _{v}^{a}=E_{v}^{a} \).
Its divergence becomes 
\[
\nabla _{a}k^{a}=\omega ^{a}\, _{va}=\frac{2g'}{gA}=-\frac{u}{2mr}\, \, \textrm{and therefore}\, \, \textrm{ }(\bar{k}\nabla )(\nabla k)=\left( \frac{2g'}{gA}\right) ^{.}=-\frac{1}{4m^{2}},\]
 where the conjugate null-direction \( \bar{k}^{a}=E^{a}_{u}=\partial _{u}^{a} \)
is uniquely defined due to spherical symmetry. Putting everything together one
finds
\[
-\frac{1}{8m^{2}}(\Delta _{S^{2}}-1)\tilde{f}(\theta ,\phi )=8\pi p\delta _{N}^{(2)},\]
which is precisely the \( \text {'t\, Hooft} \)-Dray equation for the reduced
profile without encountering any undefined \( \delta ^{2} \) expressions. Let
us finally note in passing that the results for the AS-geometry are trivially
reproduced in the proposed framework. Since the background geometry is flat
space and the null hypersurface is taken to be a null hyperplane (\ref{tHDeq})
becomes the well-known equation for the AS-profile, i.e.
\[
-\frac{1}{2}\Delta _{R^{2}}\tilde{f}=8\pi p\delta ^{(2)}(x).\]

\section*{\large\bit Conclusion}

In the present work we considered geometries belonging to the generalized Kerr-Schild
class in order to describe the change in the gravitational field generated by
a particle moving on the horizon of a stationary black hole. Since the horizon
is a null hypersurface one can more generally ask the question what kind of
restrictions such a ``localized'' change in the gravitational field imposes
on the background geometry of the Kerr-Schild decomposition. As might have been
expected those restrictions manifest themself in restrictions on the null-congruence
which forms the horizon and in particular requires the geometry to be algebraically
special along the null surface in both its Weyl and Ricci tensors. Under these
conditions we were able to derive the generalized \( \text {'t\, Hooft} \)
equation for the reduced profile function \( \tilde{f} \). Since our formalism
is general, i.e. does not rely on spherical symmetry in particular, it would
interesting to apply it to a rotating, i.e. Kerr black hole. Work in this direction
is currently in progress.

{\par\noindent \raggedright \vfill\par}

{\par\noindent \raggedright \textbf{Acknowledgment:} The author would like to
thank to Werner Israel for discussions on the generalized Kerr Schild class
which actually led to the present paper and Peter Aichelburg for reading part
of the manuscript.\par}

\end{document}